\documentclass[nofootinbib]{revtex4-2} 
\usepackage[utf8]{inputenc}
\usepackage{longtable}
\usepackage{xcolor}
\usepackage{blindtext,alltt}
\usepackage{amsmath}
\usepackage{mathrsfs}
\usepackage{amsfonts}
\usepackage{amssymb}
\usepackage{soul}
\usepackage{graphicx}
\usepackage{tikz}
\usepackage{bm}
\usetikzlibrary{cd, positioning}

\usepackage{xcolor}

\usepackage{ulem}
\usepackage{slashed}
\newcommand{\be}{\begin{equation}}
\newcommand{\ee}{\end{equation}}
\newcommand{\bear}{\begin{eqnarray}}
\newcommand{\ear}{\end{eqnarray}}
\newcommand{\half}{\frac{1}{2}}
\newcommand{\nn}{\nonumber}

\begin{document}
\title{New asymptotic techniques for the partial-wave-cutoff method for calculating the QED one loop effective action}

\author{Adolfo Huet}
\email{Corresponding author: adolfo.huet@uaq.mx}
\affiliation{Facultad de Ingenier\'{\i}a, Universidad Aut\'onoma de Quer\'etaro, Cerro de las Campanas s/n, Colonia Las Campanas, Centro Universitario, 76010, Quer\'etaro, Quer\'etaro, M\'exico.}

\author{Idrish Huet}
\email{idrish.huet@gmail.com}
\affiliation{Facultad de Ciencias en F\'isica y Matem\'aticas,\\Universidad Aut\'onoma de Chiapas,\\ Ciudad Universitaria, Tuxtla Guti\'errez 29050, M\'exico.}

\author{Octavio Cornejo-P\'erez}
\email{octavio.cornejo@uaq.mx}
\affiliation{Facultad de Ingenier\'{\i}a, Universidad Aut\'onoma de Quer\'etaro, Cerro de las Campanas s/n, Colonia Las Campanas, Centro Universitario, 76010, Quer\'etaro, Quer\'etaro, M\'exico.
octavio.cornejo@uaq.mx}


\begin{abstract}
The Gel'fand-Yaglom theorem has been used to calculate the one-loop effective action in quantum field theory by means of the ``partial-wave-cutoff method". This method works well for a wide class of background fields and is essentially exact. However, its implementation has been semi-analytical so far since it involves solving a non-linear ordinary differential equation for which solutions are in general unknown. Within the context of quantum electrodynamics (QED) and $O(2)\times O(3)$ symmetric backgrounds,  we present two complementary asymptotic methods that provide approximate analytical solutions to this equation. We test these approximations for different background field configurations and mass regimes and demonstrate that the effective action can indeed be calculated with good accuracy using these asymptotic expressions. To further probe these methods, we analyze the massless limit of the effective action and obtain its divergence structure with respect to the radial suppression parameter of the background field, comparing our findings with previously reported results. 
\end{abstract}
\maketitle

\section{Introduction}

The Gel'fand-Yaglom Theorem (GYT) \cite{gel1960integration} has inspired new methods for non-perturbative QFT \cite{dunne2008functional,ttira2011lifshitz}. In particular, it has been applied to the calculation of the one-loop effective action in QED. The effective action is an essential quantity in quantum field theory that provides important information on the interaction of quantum and classical fields. It is by its nature a non-perturbative quantity whose calculation is challenging and analytically solvable cases are very limited. In QED, for instance, there are only a few background field configurations that allow for an analytical calculation of the one-loop effective action \cite{dunne2005fields} and, in general, numerical methods are required, see for instance, the worldline montecarlo method \cite{gies2005quantum,gies2001quantum}. 

It is therefore of interest to explore and develop paths that lead to analytical approximations of the effective action. To this end some approximate methods are already well-known, for example, an analytical approach is possible when the mass of the particle is large since this limit is consistent with the heat kernel expansion, then the terms of this expansion (that involves inverse powers of the mass), can be systematically computed \cite{novikov1984calculations, ball1989chiral, schmidt1993calculation}. This method is straight forward but it has the important limitation that it can't be applied for small values of the mass. Within this mass regime, in the context of the QCD instanton, a small mass expansion has been carried out \cite{hur2009analytic}. There are also some additional approximation methods such as the derivative expansion \cite{DerivativeExpansions,PhysRevD.51.R2513,doi:10.1139/p96-044,salcedo2001derivative,PhysRevD.83.105013} that expands around the soluble constant field strength system. Finally, the semiclassical worldline instantons method has been used to calculate the imaginary part of the one-loop  QED effective action for a class of inhomogeneous backgrounds\cite{affleck1982pair,dunne2005worldline}. 

In this work we focus on a powerful semi-analytical method based on the GYT which can be used for calculating the spinor/scalar effective one-loop effective action for a wide class of backgrounds without resorting  to the aforementioned approximations. The method relies on the fact that the one-loop effective action is directly obtained from the spinor/scalar determinant \cite{PhysRevD.14.3432,PhysRevD.16.417,PhysRevD.16.1052,PhysRevD.17.3238}, and thus, the GYT makes it possible to calculate such determinants in a simpler and more direct way. For instance, this method allows the calculation of the spinor or scalar determinant when the background field is radially symmetric in the euclidean space and the background field is such that it permits a decomposition of the full determinant into partial-wave one-dimensional determinants of definite angular momentum  \cite{PhysRevLett.94.072001, PhysRevD.87.125020,PhysRevD.74.085025,PhysRevD.77.045004,PhysRevD.77.125033}. This procedure was called the ``partial-wave-cutoff method" in \cite{PhysRevD.74.085025} because the partial wave operators are classified into low and high angular momentum modes. The ``high-modes" can be treated analytically with a WKB expansion of partial amplitudes and thus, one can perform the required renormalization subtractions while the GYT is used for the calculation of the ``low-modes'' sector. The calculation of those low partial-wave determinants amounts to solving the related ordinary differential equation (ODE) and obtain the limiting value of those solutions at $r \rightarrow \infty$.

The reason why this part of the calculation has been performed numerically within this context, is that the partial-wave ODE happens to be non-linear and exact solutions are unknown for most background fields. Therefore, the low-modes-equation is the reason why this method is not entirely analytical. This fact brings along some disadvantages; first we lose information by having to deal with tables of numerical data instead of analytical expressions and second, depending on the specifics of the background field, the computational work involved can become cumbersome and impractical. 

However, it should be noted that complete knowledge of the solution of the related ODE is not required, in fact we only need the limiting value of the solution as $r \rightarrow \infty$. This raises the question of how precise are some of the techniques of asymptotic analysis in order to predict this limiting value, and how reliable those approximations would be for different background fields or mass regimes.

In this paper, we present two complementary methods that are based on a classical technique of asymptotic analysis for systematically expanding the low-mode equation so that one can express the solution as an integral expression that approximates well the exact solution and allows the calculation of the needed limiting values. These integral expressions can be considered approximate quadratures for the non-linear partial wave equation for which the exact analytical solution is unknown.

We test these methods for a concrete example, namely, the calculation of the one-loop spinor/scalar effective action under the background of an $O(2) \times O(3)$ symmetric field with a profile that was recently studied \cite{PhysRevD.87.125020}. Although we will work with a specific background profile to obtain numerical information and for specific analysis, our main results will be given in terms of a profile function $g(r)$ (in euclidean space). This function is only required to have an appropriate fall-off rate, but apart from that, it is essentially arbitrary. 

In section \ref{PWC} we present a summary of the partial-wave-cutoff method as applied to the class of $O(2) \times O(3)$ considered. In section \ref{Asymptotics}, two asymptotic methods are applied to approximate the solutions of the partial-wave ODE. The corresponding expressions for the one-loop effective action are obtained and numerical evidence of the accuracy of the new methods is presented. In section \ref{Scope} we analyze the expressions obtained for the partial-wave solutions and effective action on the massless limit to test some of their scope and limitations. Finally, in section \ref{conclusions} we present our conclusions.

\section{The partial-wave-cutoff method \label{PWC}}
A system that illustrates many important features of the GYT based techniques for QFT, is the problem of the QED one-loop effective action under the following background field:
\begin{eqnarray} \label{Apotential}
A_\mu (x) &=& \eta_{\mu \nu}^3 x_\nu g(r)\;,
\end{eqnarray}
where $\eta_{\mu \nu}^3$ are the t'Hooft symbols and $g(r)$ is a radial profile function in euclidean space \cite{PhysRevD.14.3432}.
This type of field has an $O(2) \times O(3)$ symmetry that leads to the remarkable property that it makes the Klein-Gordon and Dirac operators isospectral except for possible zero modes\cite{PhysRevD.6.3445,PhysRevD.10.2399,bogomol1981asymptotic,bogomol1982asymptotic,PhysRevD.75.065002,PhysRevD.81.107701}, this allows one to obtain both the spinor and scalar actions from almost the same calculation. To briefly summarize  the method we will present the equations used  for the spinor action, following ref.\cite{PhysRevD.87.125020}. The one-loop spinor effective action (in euclidean space) is given by
\begin{eqnarray}
\Gamma[A] =-{\rm ln \;det} (-{\slashed D}+m)= -\frac{1}{2}{\rm ln \;det} (-{\slashed D}^2+m^2) \;,
\end{eqnarray}
where ${\slashed D}=\gamma_\mu (\partial_\mu + ie A_\mu (x))$, we set $e=1$. As it has been demonstrated, the chiral decomposition for this type of field leads to $\Gamma[A]=\Gamma[A]^{(+)}+\Gamma[A]^{(-)}$ where
\begin{eqnarray}
\Gamma^{(+)}[A] &\equiv& -\frac{1}{2}{\rm ln \;det} (-D^2+m^2+ \frac{1}{2} F_{\mu \nu} \Bar{\eta}^{a}_{\mu \nu} \sigma_a) \\
\Gamma^{(-)}[A] &\equiv& -\frac{1}{2}{\rm ln \;det} (-D^2+m^2+ \frac{1}{2} F_{\mu \nu} \eta^{a}_{\mu \nu} \sigma_a)
\end{eqnarray}
From this we can write $\Gamma[A]=2 \Gamma[A]^{(\pm)} \mp (\Gamma[A]^{(+)}-\Gamma[A]^{(-)}) \equiv 2 \Gamma[A]^{(\pm)} \mp \Delta \Gamma[A]$. It is known that after renormalization one obtains  
\be
\Delta \Gamma_{\rm ren} [A] = \frac{1}{2} \frac{1}{(4 \pi)^2} \ln \left( \frac{m^2}{\mu^2} \right) \int dx^4 F_{\mu \nu} \Tilde{F}_{\mu \nu}
\ee
and that non zero $\Delta \Gamma_{\rm ren} [A]$ would indicate the presence of anomalous zero-modes.  For all our numerical tests we work with a previously studied \cite{spinorDunne, PhysRevD.87.125020} profile function:
 \be
 g(r) = \frac{\nu e^{-\alpha r^2}}{\rho^2   + r^2} \;.
 \label{gdef}
 \ee
 where $\nu, \alpha$ and $\rho$ are positive constants.
The first two constants are related to the field strength, and the fall-off rate respectively and we set $\rho=1$ for all our numerical calculations. As it has been shown \cite{PhysRevD.87.125020,spinorDunne}, for $\alpha >0$ implies $\Delta \Gamma_{\rm ren} [A]=0$ and one may obtain the the spinor effective action from the calculation of one chirality sector, namely  $\Gamma_{{\rm  ren}}^{{\rm sp}}[A] = 2 \Gamma_{{\rm ren}}^{(-)}[A]=-2 \Gamma_{{\rm ren}}^{{\rm sc}}[A]$. We will exploit this relation for efficiency in numerical calculations. However, our main results are derived for an arbitrary profile  funxtion $g(r)$.

Using the partial-wave decomposition that this background allows, we obtain the spinor effective action as a sum over definite angular momentum determinants:
\begin{eqnarray}
\Gamma &=&  -\sum_{s=\pm} \sum_{l =
0, \frac{1}{2} , 1, \ldots}^\infty \; \Omega(l) \sum_{l_3 = -l}^{l}
\ln \left( \frac{ {\rm{det}}(m^2+\mathcal{H}_{(l, l_3, s)} )}{{\rm{det}}(m^2+\mathcal{H}_{(l, l_3, s)}^{{\rm{free}}}
)} \right) \, ,
\label{GammaPartialwaves}
\end{eqnarray}
where $\Omega(l) = 2 l +1$ and 
\begin{equation} \label{sc-decomp}
m^2+\mathcal{H}_{(l, l_3, s)}= - \left[ \partial_r^2  
+\frac{4 (l+3)}{r^2}\partial_r - r^2 g(r)^2
-4 g(r) l_3 -m^2 \mp (4 g(r)+r g'(r)) \right] \, .
\end{equation}
The determinants of these partial wave operators can be calculated using the GYT, however, the problem of calculating the full effective action presents an additional challenge because the simple addition of all the partial wave determinants neglects renormalization and turns out to be divergent. The partial-wave-cutoff method consists of splitting the sum in two parts by setting an arbitrary angular momentum cutoff $L$, then we have:
\be
 \Gamma =  -\sum_{s=\pm} \sum_{l =
0, \frac{1}{2} , 1, \ldots}^L \; \Omega(l) \sum_{l_3 = -l}^{l}
\ln \left( \frac{ {\rm{det}}(m^2+\mathcal{H}_{(l, l_3, s)} )}{{\rm{det}}(m^2+\mathcal{H}_{(l, l_3, s)}^{{\rm{free}}}
)} \right) 
-\sum_{s=\pm} \sum_{l =
L, L+\frac{1}{2} , \ldots}^\infty \; \Omega(l) \sum_{l_3 = -l}^{l}
\ln \left( \frac{ {\rm{det}}(m^2+\mathcal{H}_{(l, l_3, s)} )}{{\rm{det}}(m^2+\mathcal{H}_{(l, l_3, s)}^{{\rm{free}}}
)} \right)
\ee
The first part has the ``low-modes" that run from $l=0$ to $l=L$, it is for this sum that we will apply the GYT. The second part has the ``high-modes" $l\geq L$. This latter sum is divergent but as the cutoff $L$ becomes larger  the partial-wave determinants are well approximated by a WKB expansion, this produces simpler expressions allowing each sum to be performed. For $l$ the Euler-Maclaurin summation technique is used. After this, one has an explicit analytical result from which the divergences can be identified an removed with dimensional regularization (MS) by means of the counterterm $\delta\Gamma$. Therefore, we have:
\begin{equation}
\Gamma = \Gamma_{\rm L} + \Gamma_{\rm H} + \delta\Gamma = \Gamma_{\rm L} + \Gamma_{\rm {H, ren}}
\label{GammaHLdelta}
\end{equation}
The final expressions for the renormalized high-mode sum for both the spinor and scalar cases are provided in \ref{highmodes}.

While (\ref{GammaHLdelta}) explicitly depends on the arbitrary cutoff, one finds that as $L \to \infty$ the effective action converges well to fixed value.  We remark the GYT is not needed for the high-modes but only for the low-modes where the WKB expansion doesn't work. The GYT is applied to the calculation of the low-modes as follows; let $\mathcal{M}_1$ and we $\mathcal{M}_2$ be the two differential operators that each define a corresponding initial value problem: \be
\mathcal{M}_i \Phi_i =0, \quad  \Phi(r) \overset{r \to 0}{\sim} r^{2l}
\ee 
The ratio of the determinants is given by 

\be
 \frac{{\rm det} \mathcal{M}_1}{{\rm det} \mathcal{M}_2}  = \lim_{R \to \infty} \left( \frac{\Phi_1(R)}{\Phi_2(R)}  \right)\;. \label{GYratio}
\ee

For the system at hand, the specific equation to solve is the following
\be
\Phi_{\pm}^{''}(r)+\frac{4l+3}{r}\Phi_{\pm}^{'}(r)-(m^2+ 4l_3 g(r)+r^2 g(r)^2\mp h(r))\Phi_{\pm}(r) = 0\;,
\ee
where $h(r) \equiv 4g(r)+rg'(r)$. The corresponding equation for the scalar case is obtained by setting $h(r)=0$.
For this case, the ratio in (\ref{GYratio}) is to be taken with respect to the free equation for which $g(r)=0$. 

It is convenient to introduce the function 
\be
S(r)= {\rm ln} \left( \frac{\Phi(r)}{\Phi_{{\rm free}}(r)}\right)
\ee
which is numerically easier to handle. Exploiting the useful fact that $\Phi_{{\rm free}}(r)$ is known exactly, we end up with a more tractable problem:

\bear
	\frac{d^2 S_{\pm}^{l,l_3} }{dr^2} 
	+ \left( \frac{d S_{\pm}^{l,l_3} }{dr}  \right)^2 
	+ W_l (m,r) \frac{d S_{\pm}^{l,l_3} }{dr}  \label{Smodes}
	&=& V_{l_3}^{\pm} (r) \;, \nn \\
	S_{\pm}^{l,l_3}(0)&=&0 \; , \nn  \\	
	\frac{d S_{\pm}^{l,l_3} }{dr} (0) &=& 0 \; ,
\label{Seq}
\ear 
where
\bear
	W_l (m,r) &\equiv& \frac{1}{r} 
		+ 2m \frac{I_{2 l +1}^{'} (m r)}{I_{2 l +1}(m r)} \; , \\ \label{Vofg}
	V_{l_3}^{\pm} (r) &\equiv& 4 l_3 g(r)  + r^2 g(r)^2 \mp h(r) \; ,	
\ear
and $I_{a}(x)$ is the modified Bessel function of the first kind. Applying the GYT, the partial wave determinants are obtained as the asymptotic limit of the solution of equation (\ref{Seq}):
 \be
 \ln \left( \frac{ {\rm{det}}(m^2+\mathcal{H}_{(l, l_3, \pm)} )}{{\rm{det}}(m^2+\mathcal{H}_{(l, l_3, \pm)}^{{\rm{free}}}
)} \right) = \lim_{r \to \infty} S_{\pm}^{l,l_3}(r)
 \ee
 So far, in previous works using the partial-wave-cutoff method, such as \cite{PhysRevD.77.045004,PhysRevLett.94.072001, PhysRevD.83.105013, PhysRevD.87.125020}, equation (\ref{Seq}) has been solved numerically. However, a full knowledge of $S_{\pm}^{l,l_3}(r)$ is not required, since the determinant is given by its asymptotic value at infinity. 
 
 In the next section we apply asymptotic analysis  techniques to obtain approximations to $S_{\pm}^{l,l_3}(\infty)$ in the form of integral expressions that apply to a wide class of background fields. We will test those approximations for different field configurations and mass regimes and use them calculate the one-loop effective action.
 
\section{Asymptotic methods for the partial wave equation \label{Asymptotics}}

\subsection{Dominant balance method}

We shall investigate an approximate solution to the partial wave equations  based on the dominant balance (DB) method \cite{bender2013advanced}, 
to this end let us first rewrite  (\ref{Seq}) as a first order Ricatti equation by introducing the dimensionless variable $z = mr$ and the shorthand $\Sigma (z) \equiv \frac{d S_{\pm}^{l,l_3} (z)}{dz}$,

\begin{equation} \label{Ricatti}
    \frac{d\Sigma (z)}{dz}  +  \Sigma^2 (z) + w(z) \Sigma(z) = v (z), \quad \Sigma(0) = 0,
\end{equation}
where indices have been omitted as a shorthand and we have defined the dimensionless functions $w (z) = \frac{1}{z} + 2 \frac{I'_{2l+1}(z)}{I_{2l+1}(z)}$ and $v(z) = \frac{1}{m^2}V_{l_3}^{\pm} (r)$.

We introduce a deformation parameter $\epsilon$ in (\ref{Ricatti}) such manner that the new Riccati equation is easily solvable for $\epsilon =0$ and also recovers its original form for $\epsilon =1$, namely

\begin{equation} \label{BalDom}
    \frac{d\Sigma (z,\epsilon)}{dz}  + \epsilon \Sigma^2 (z,\epsilon) + w(z) \Sigma(z,\epsilon) = v (z), \quad \Sigma(0,\epsilon) = 0,
\end{equation}
where the following formal series ansatz is used
\be \label{Sigma}
\Sigma(z,\epsilon) = \sum_{n=0}^{\infty} \epsilon^n \Sigma_{n} (z).
\ee
After substitution, we obtain an iterative sequence of linear first order equations 

\bear
\frac{d\Sigma_0 (z)}{dz} + w(z) \Sigma_0 (z) &=& v(z) \label{sigma0} \\
\frac{d \Sigma_n (z)}{dz} + w(z) \Sigma_n (z) &=& - \sum_{i+j = n-1} \Sigma_i (z) \Sigma_j (z), \qquad n \geq 1
\ear
each of which can be integrated straightforwardly\footnote{Notice the integrating factor $\exp \left( \int_0^z w(u) \,du \right) = z I_{2l+1}^2 (z)$ renders the equations exact.}. Then, the $n$-th inhomogeneous equation for $n \geq 1$ is solved as

\begin{equation} \label{Sigman}
   \Sigma_n (z) = - \frac{1}{z I_{2l+1}^2 (z)} \sum_{i+j = n-1} \int_{0}^{z}  y I_{2l+1}^2 (y) \Sigma_i (y) \Sigma_{j}(y)~ dy
\end{equation}
For instance, the first level of approximation ($n=0$) yields
\begin{equation} \label{Sigma0}
  \Sigma_0 (z) = \frac{1}{z I_{2l+1}^2 (z)} \int_0^z    y I_{2l+1}^2 (y)  v(y) ~dy
\end{equation}
Its important to remark that the boundary condition in (\ref{BalDom}) (required by the GYT) has already been taken into account in (\ref{Sigman})  and (\ref{Sigma0})  implying that $\Sigma_n (0) = 0$.

\begin{figure}
  \centering
  \includegraphics[width=.55\linewidth]{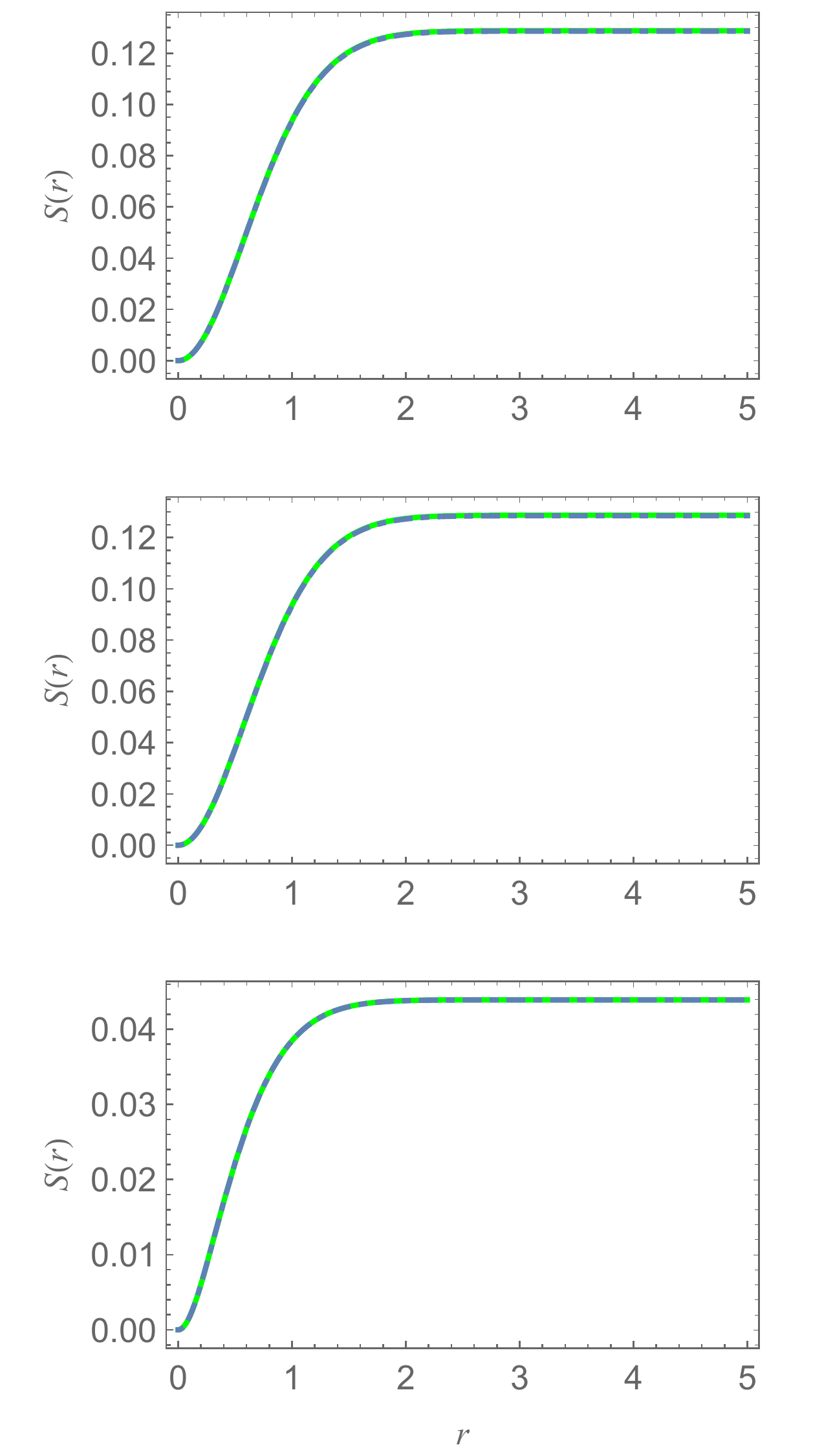}
\caption{ Plots of $S_{+}^{l,l_3}(r)$ (dashed line) and $S_0(r)$ (green line) for $\nu=1$, $\alpha=1$, $\rho=1$, $l=10$, $l_3=5$ and $m=1/1000$ (upper graph), $m=1$ (middle graph), $m=100$ (bottom graph).}
\label{figure1}
\end{figure}

The question of whether or not the actual solution to (\ref{Ricatti}) will be obtained after setting $\epsilon =1$ in (\ref{Sigma}), that is whether $\Sigma (z) = \Sigma (z,1)$, has an affirmative answer if the convergence radius for the power series is $\rho >1$. This question has to be investigated on a case by case basis since it depends on the background field. For the problem at hand, we have found that taking a large enough number of terms in the series does lead to a convergent solution. In the dominant balance method, the key is to identify which term in (\ref{BalDom}) will be dominant by the others, then we can set the parameter $\epsilon=1$ which serves as a label of the order of subdominance of different contributions. From this point of view in this instance the method of dominant balance can be conceived as a perturbative method around the $\epsilon =0$ solution. A quantitative criterion for fast convergence of the method is given by

\begin{equation}
    \sup \{ |v(z)/w^2(z)| \,: \,z \geq 0 \} \ll 1.
\label{DBcriterion}
\end{equation}
Taking into account the boundary condition $S_{\pm}^{l,l_3}(0) = 0$ we obtain for each partial wave
\be \label{asymp}
S_{\pm}^{l,l_3}(z) = \sum_{n=0}^{\infty} S_{n} (z),\quad\mbox{where}\quad S_{n} (z) :=  \int_0^z  \Sigma_n (u) ~ du
\ee
From equations (\ref{Sigma0}) and (\ref{asymp}) we get the following expression for the leading contribution in the approximation:
\be \label{s0first}
S_{0}(z)= \int_0^z dx \frac{1}{x I_{2l+1}^2 (x)} \int_0^x  dy ~v(y)~ y I_{2l+1}^2 (y); 
\ee
A series of transformations, detailed in appendix \ref{A}, reduce the number of integrations required, in particular (\ref{s0first}) can be reduced to a single integral:

\be \label{S0}
S_{0}(z) = \int_0^{z} dy ~v(y) y \frac{I_{2l+1} (y)}{I_{2l+1}(z)}   \left(K_{2l+1}(y) I_{2l+1}(z) -K_{2l+1}(z) I_{2l+1}(y) \right) \;.
\ee
As shown in fig.\ref{figure1} this basic approximation already shows good agreement with the exact numerically obtained solution, for either small or large mass.

The next order of approximation yields
\be 
\Sigma_1 (z) =  -\frac{1}{z I_{2l+1}^2 (z)} \int_0^z  dx ~\frac{1}{x I_{2l+1}^2 (x)} \left( \int_0^x dy ~I_{2l+1}^2 (y) v(y) y \right)^2
\ee
which leads to 
\be \label{S1direct}
S_1 (z) = - \int_0^z dt \int_0^z dx \frac{1}{tI_{2l+1}^2 (t) x I_{2l+1}^2 (x)} \left(\int_{0}^{x} dy ~I^2_{2l+1}(y) v(y) y \right)^2
\ee
As shown in \ref{A}, it is remarkable that this expression can be reduced to a double integration
\bear \label{S1}
S_1 (z) &=& -\half \left(\frac{K_{2l+1}(z)}{I_{2l+1}(z)} \right)^2 \left( \int_0^{z} dx ~I_{2l+1}^2(x) v(x) x \right)^2  \\
&+&  2 \frac{K_{2l+1}(z)}{I_{2l+1}(z)}  \int_0^{z} dy~ K_{2l+1}(y) I_{2l+1} (y) v(y)y \int_{0}^y dx~ I_{2l+1}^2 (x) v(x) x \nn \\
& -&  \int_0^z dy ~K_{2l+1}^2(y) v(y) y 
\int_0^y dx ~I_{2l+1}^2 (x) v(x) x  \nn
\ear
In addition to reducing the amount of integrations, the main advantage of expressions (\ref{S0},\ref{S1}) over  (\ref{s0first},\ref{S1direct}) is that they take a much simpler form  at $z \to \infty$ which is the relevant quantity in our problem, a particularly simple formula is obtained:

\bear \label{Sseries}
S_{\pm}^{l,l_3}(\infty)&=& \int_{0}^{\infty}dy~ K_{2l+1}(y) I_{2l+1} (y) v(y) y  \\
&-&  \int_0^\infty dy~ K^2_{2l+1} (y) v(y) y \int_0^y dx~ I_{2l+1} (x) v(x) x \,\,+\,\, \cdots \nn
\ear

Although we shall stop at two terms, one can treat in a similar fashion the $n$-th order correction, starting off from (\ref{Sigman}) and integrating by parts as in \ref{A}, using the identity (\ref{derbessel}), to obtain the general expression 

\begin{equation}
S_{n} (\infty) = -\sum_{i+j = n-1}\int_0^{\infty} K_{2l+1}(y)I_{2l+1}(y)  \Sigma_{i} (y) \Sigma_j (y) ~dy
\end{equation}

Using this approximation, we have been able to reproduce results obtained with, the previous angular-momentum cut-off methodology for the full effective action \cite{PhysRevD.87.125020}. From the approximations derived above one arrives to the following expression
\be
\Gamma_{{\rm L}} =  -\sum_{l =
0, \frac{1}{2} , 1, \ldots}^L  (2l+1) \sum_{l_3 = -l}^{l} (S^{l, l_3}_{-}(\infty)+ S^{l, l_3}_{+}(\infty))
\label{GammaGYT}
\ee
After substituting $S_0(r)$ and $S_1(r)$ it is possible to calculate the sum over $l_3$ and after some simplifications we obtain  $\Gamma_{\rm L} \approx \Gamma_0 + \Gamma_1$, where
\be
\Gamma_0 =  -\sum_{l =
0, \frac{1}{2} , 1, \ldots}^L 2(2l+1)^2  \int_0^\infty K_{2l+1}(mr) I_{2l+1}(mr) r^3 g^2(r) dr
\ee
and
\begin{eqnarray}
\Gamma_1 &=&  \sum_{l =
0, \frac{1}{2} , 1, \ldots}^L \frac{32}{3} (2l+1)^2 l(l+1) \int_0^\infty d\rho
\frac{K_{2l+1}(m\rho)}{\rho I_{2l+1}^3(m\rho)} \left( \int_0^\rho dr I_{2l+1}^2(mr) r g(r) \right)^2 \nonumber \\
&+&\sum_{l =
0, \frac{1}{2} , 1, \ldots}^L 2(2l+1)^2 \int_0^\infty d\rho
\frac{K_{2l+1}(m\rho)}{\rho I_{2l+1}^3(m\rho)} \left( \int_0^\rho dr I_{2l+1}^2(mr) r^3 g(r)^2 \right)^2 
\end{eqnarray}
As indicated on section \ref{PWC},  renormalization is carried out in the high-mode contribution $\Gamma_{{\rm H, ren}}$ by means of  the known procedure. After accounting for both parts, the effective action is then approximated as:
\be
\Gamma_{{\rm ren}} \approx \Gamma_1+\Gamma_0+\Gamma_{{\rm H, ren}}
\label{GammaDB}
\ee
 What has been achieved, is an analytical expression for the low-modes contribution that in previous works was calculated by means of numerical solutions to equation (\ref{Seq}), while also reducing the double sum to a single sum. We emphasize that the angular-momentum cutoff method represents a particularly hard test on the approximation method, not only because we are accumulating approximation error from each partial wave, but also because both $\Gamma_{\rm L}$ and  $\Gamma_{{\rm H, ren}}$ contain divergences of the type $\ln L$, $L$, and $L^2$ that need to exactly cancel as $L \to \infty$, leaving a finite value for the effective action.

\begin{figure}
  \centering
  \includegraphics[width=.55\linewidth]{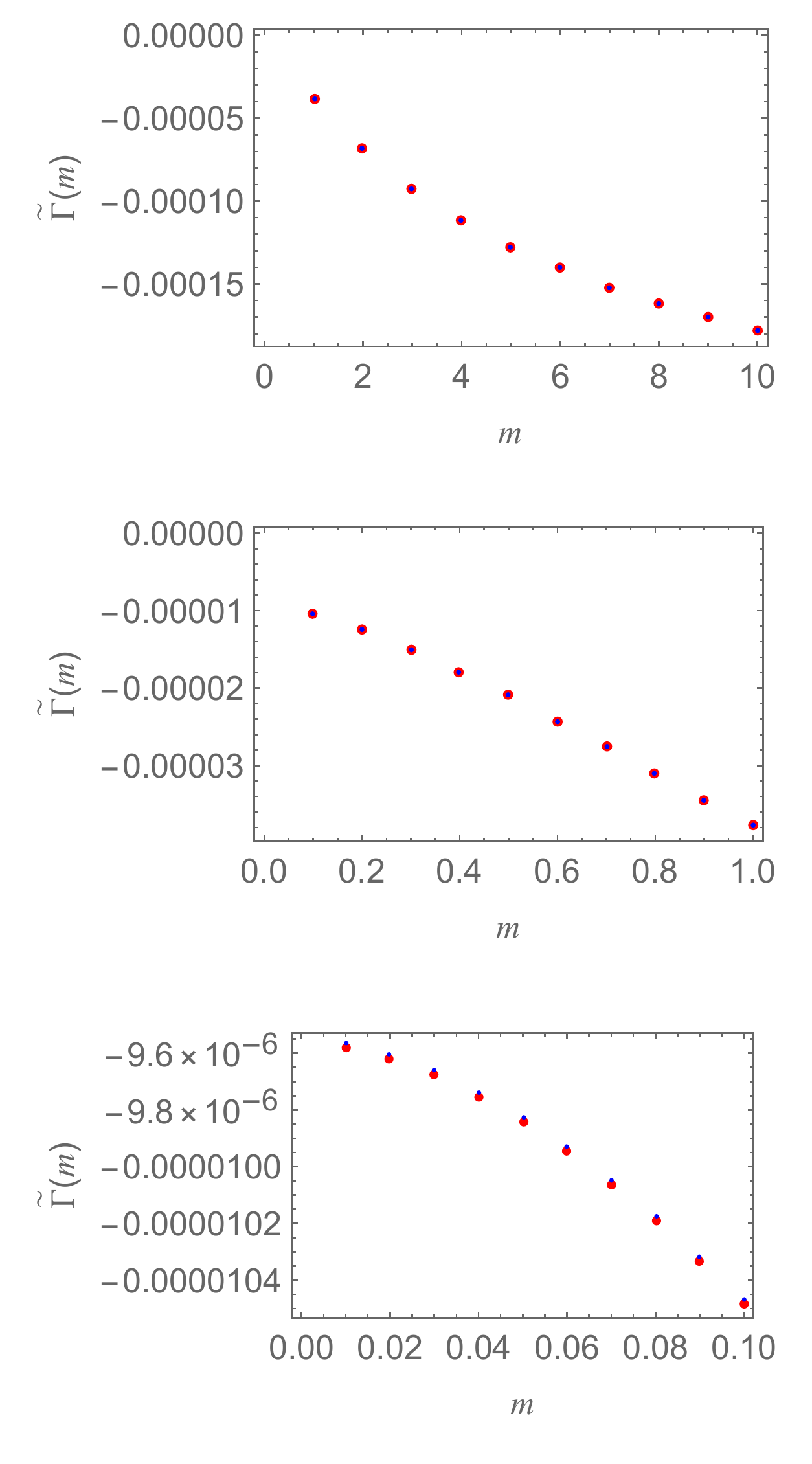}
\caption{ Plots of $\Tilde{\Gamma}(m)$ for $\nu=1/10$, $\alpha=2$, $\rho=1$. 
The red (big) dots correspond to DB method while the blue (small) dots show the numerical (exact) calculation. The upper graph shows $m=1,2, 3, \dots, 10$. The center graph shows $m=0.1,0.2, 0.3, \dots, 1$. The bottom graph shows $m=0.01,0.02, 0.03, \dots, 0.1$.}
\label{figure2}
\end{figure}

Following the type of analysis done in ref.\cite{PhysRevD.87.125020}, we point out that the on-shell renormalized effective action goes as
\be
\Gamma_{{\rm ren}}^{{\rm OS}}(m)\sim -\left( \int_0^\infty Q_{{\log}}(r) dr  \right) {\ln (m)} , \quad m \to 0\;,
\ee
where $Q_{{\log}}(r)$ is given in \ref{highmodes}.
Therefore, to present results in the small-mass regime it is convenient to define the modified effective action as
\be
\Tilde{\Gamma}_{{\rm ren}}(m) \equiv \Gamma_{{\rm ren}}^{{\rm OS}}(m) +\left( \int_0^\infty Q_{{\log}}(r) dr  \right) {\ln (m)}\;,
\ee
The modified effective action as approximated by (\ref{GammaDB}) is shown in fig. \ref{figure2} for different mass regimes. On the last graph a finer scale is used in the plot that allows us to see the small amount of error from our approximation.

\subsection{A WKB type method}
Equation (\ref{DBcriterion}) indicates the best scenario for the implementation of the DB method, however different field configurations require a different kind of asymptotic method.
Let us consider the opposite regime, that is when
\begin{equation}
     \sup \{ |v(z)/w^2(z)| \,: \,z \geq 0 \} \gg 1.
     \label{WKBcriterion}
\end{equation} 
 In such case, an asymptotic approximation of the WKB type results most useful. Let us perturbate the Riccati equation (\ref{Ricatti}) in the form

\begin{equation}
\epsilon \frac{d\sigma (z,\epsilon)}{dz} + \sigma^2 (z,\epsilon) + w(z) \sigma (z,\epsilon) = v(z), \quad \sigma(0,\epsilon)=0
\end{equation}
and propose an ansatz $\sigma (z,\epsilon)$  constrained by $\Sigma(z) = \sigma(z, 1)$ and then use an expansion of the form $\sigma (z,\epsilon) = \sum_{n=0}^{\infty} \epsilon^n \sigma_n (z)$.   
 
\begin{figure}
  \centering
  \includegraphics[width=.53\linewidth]{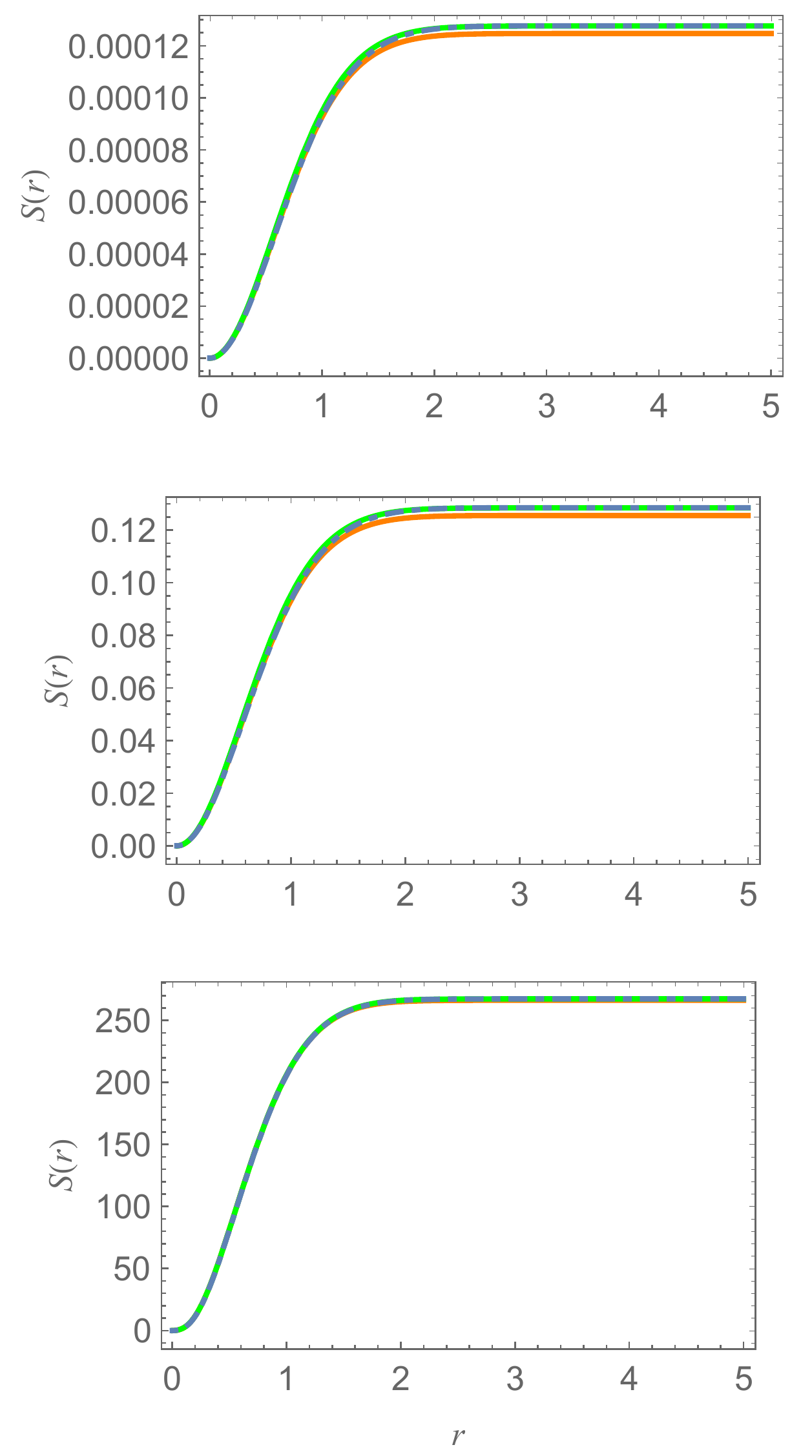}
\caption{ The plots show the exact $S^{l,l_3}_{+}(r)$ blue (dashed) line, against the first two levels of the WKB approximation: the first level as a orange (solid) line and the second level as a green (solid) line. The parameters are set to $m=1$, $\alpha=1$ and $\rho=1$. The parameter $\nu$ related to the field strength is varied: $\nu=1/100$ (upper graph), $\nu=1$ (middle graph) and $\nu=1000$  (bottom graph).}
\label{figure3}
\end{figure}

The first levels in this approximation produce the following equations:
\begin{eqnarray}
\sigma_0(z)^2 + w (z) \sigma_0 (z) - v(z) &=& 0 \\
\frac{d \sigma_0 (z)}{dz} +2 \sigma_0 (z) \sigma_1 (z) + w (z) \sigma_1 (z) &=&0 \\
 & \vdots &  \nonumber 
\end{eqnarray}

from which we obtain the counterpart to (\ref{asymp}) and (\ref{Sseries})
\be
S^{l,l_3}_{\pm} (z) = \int_0^z \sigma_0 (\zeta)  \,d\zeta +  \int_0^z  \sigma_1 (\zeta) \,d\zeta + \cdots \equiv \mathcal{S}_0 (z) + \mathcal{S}_1 (z) + \cdots
\ee
Notice that in the last expression the boundary conditions are already implemented.
An example of the performance of this approximation is shown in fig. \ref{figure3}.

The limit $r \to \infty$ provides a the following expression after integrating by parts
\bear
S^{l,l_3}_{\pm} (\infty)  &=& \int_0^{\infty} \left( \sqrt{v(z) + \frac{w^2(z)}{4}} - \frac{w(z)}{2}\right) ~dz \label{WKBpartialdet} \\ \nn 
&-& \int_0^{\infty} \left( \sqrt{v(z) + \frac{w^2(z)}{4}} - \frac{w(z)}{2}\right) \frac{2 v' + ww'}{(4 v + w^2)^{3/2}}~dz + \cdots
\ear

\begin{figure}
  \centering
  \includegraphics[width=.55\linewidth]{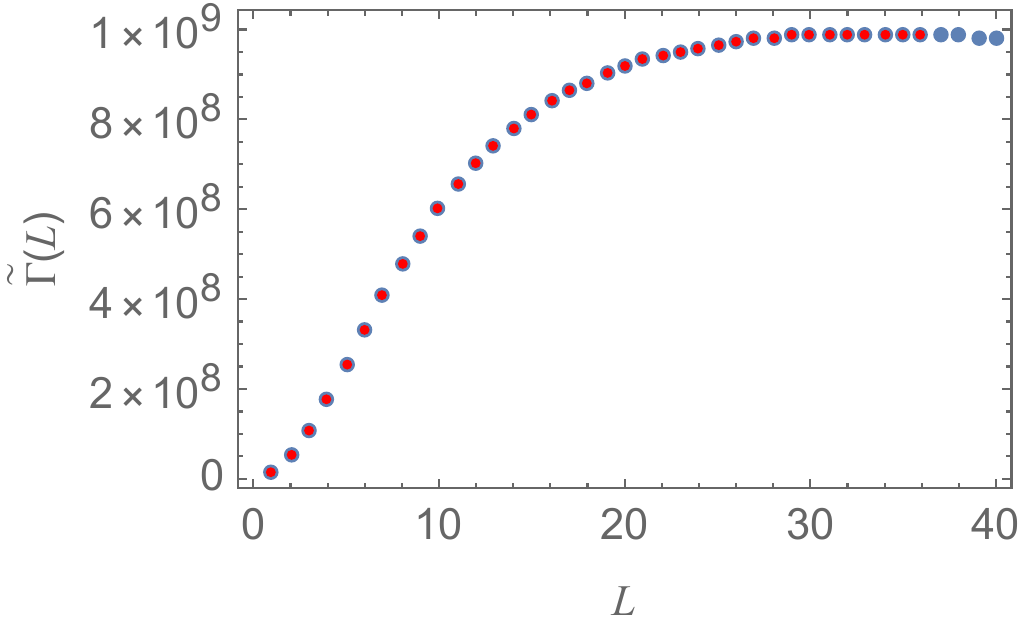}
\caption{ Plots of $\Tilde{\Gamma}(L)$ for $\nu=1000$, $\alpha=1/5$, $\rho=1$. 
The red dots correspond to the WKB type method while the blue dots represents the numerical (exact) calculation. The action is shown as a function of the angular-momentum-cutoff $L$. }
\label{figure4}
\end{figure}

\begin{figure}
  \centering
  \includegraphics[width=.55\linewidth]{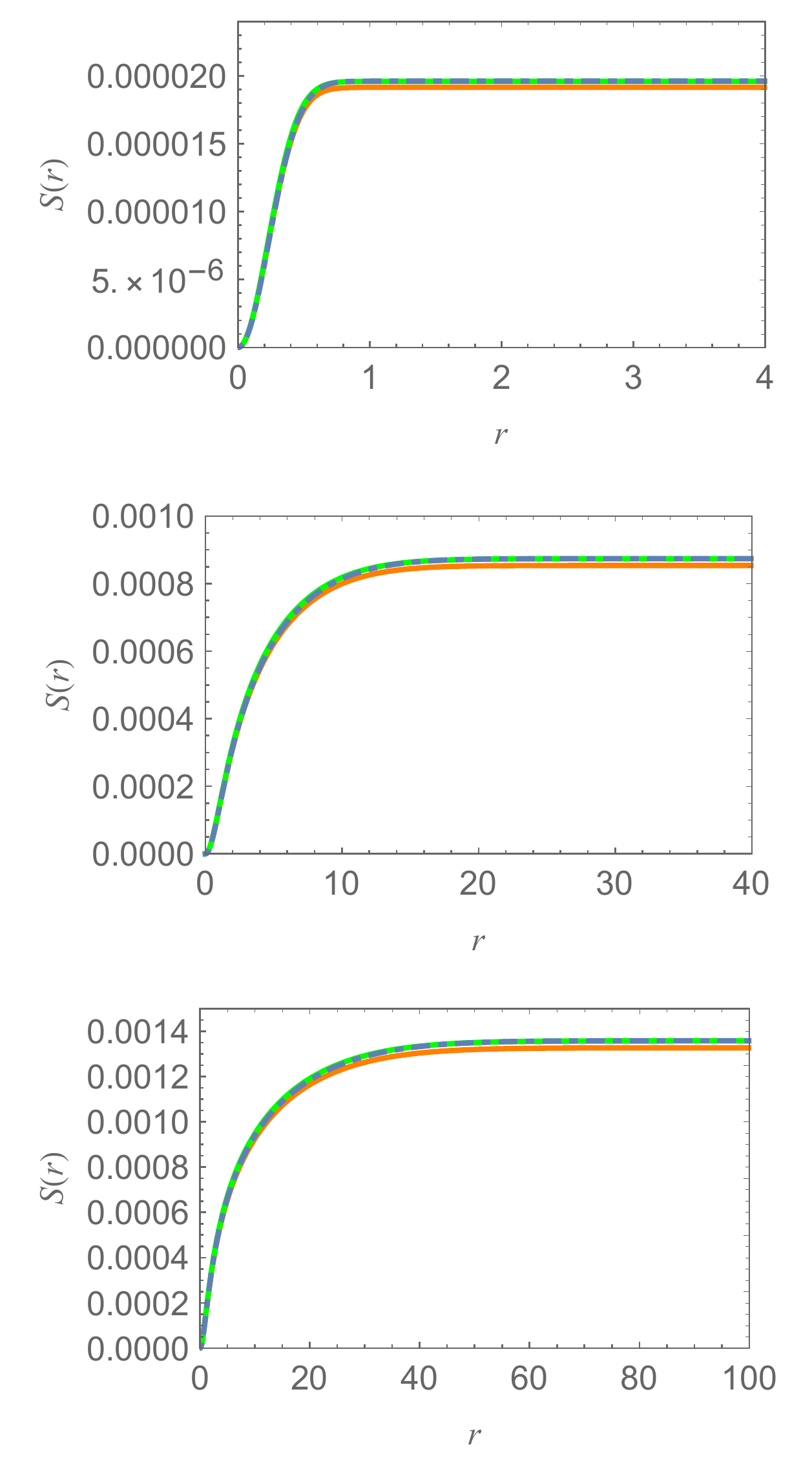}
\caption{In the plots above, we show, the exact $S^{l,l_3}_{+}(r)$ for $m=0$ as a blue (dashed) line, against the first two levels of the WKB approximation: the first level in orange and the second level as a green line. The parameters are set to be $l=10,\,l_3=5$ and $\rho=1$, while the parameter $\alpha$ related to the range of the background field is varied: $\alpha=10$ (upper graph), $\alpha=1/100$ (middle graph) and $\alpha=1/1000$  (bottom graph).}
\label{figure5}
\end{figure}

\begin{figure}
  \centering
  \includegraphics[width=.65\linewidth]{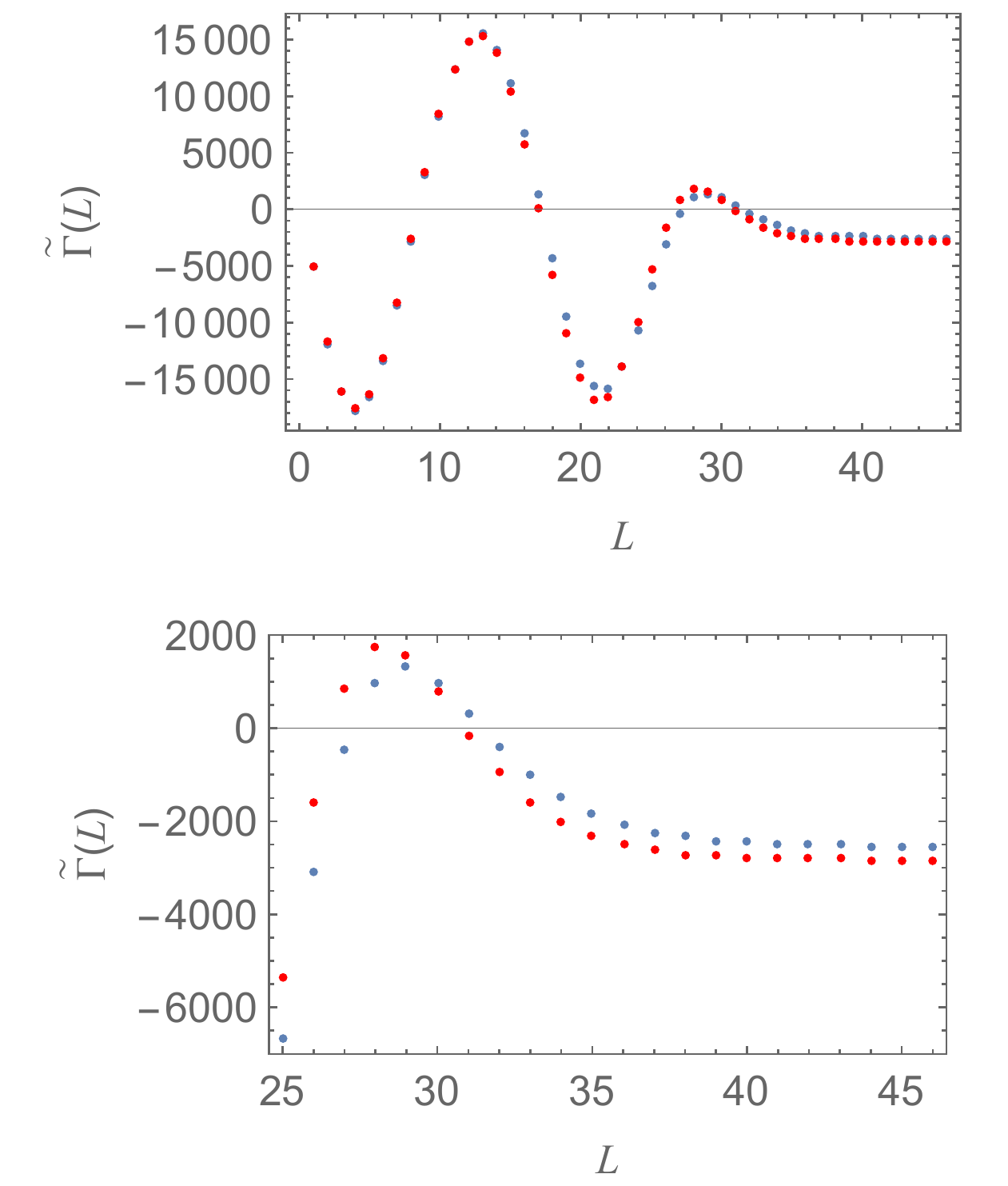}
\caption{In the plots above, we show $\Tilde{\Gamma}(L)$ for $m=0$, and $\mu=1$. The blue dots represent the WKB type method with the first two levels in the approximation, the red dots represent the corresponding exact calculation. The second (lower) plot zooms in to show the error. The parameters are set to be $\nu = 100,\,\alpha = 1/2$ and $\rho=1$. Both calculations converge for large values of $L$.}
\label{figure6}
\end{figure}

Once more, it is possible to calculate the effective actions using the partial-wave-cutoff method in conjunction with formula (\ref{WKBpartialdet}). In practice, this type of approximation may be compared with the DB method, for instance, as it is evidenced by fig. \ref{figure3} a ``strong-field background" favors the precision of the WKB type approximation. A comparison between the effective action obtained with the exact numerical solution versus the WKB type approximation is presented in fig. \ref{figure4}, we see good agreement between the exact and asymptotic methods and observe a smooth convergence for increased values of the angular momentum cutoff $L$.

Finally, we tested this approximation for the massless limit which will be discussed in the next section, as we can see in fig. (\ref{figure5}) the second level of approximation already works well. As before, it is still possible to calculate the effective action using the WKB type approximation in the massless limit. We show the convergence of $\Tilde{\Gamma}(L)$ with respect to the cutoff for both the exact and approximated low-modes. Since for this limit is impossible to have on-shell renormalization conditions,  we use $\mu=1$ as our choice for the unphysical renormalization condition.

\section{Scope and limitations  \label{Scope}}
Since the WKB type method provides analytical single integral expressions for the low-mode contribution to the effective action, it is now possible  to carry out some analysis of the partial-wave determinants and the effective action for a fixed background profile. In this section we will obtain asymptotic expressions for the small $\alpha$ regime of to probe our formulas under a second approximation. In ref.\cite{PhysRevD.87.125020} the same background profile was studied and an analysis of the massless effective action in the limit $\alpha \to 0$ was performed by means of Worldline techniques supported by the (numerical) partial-wave-cutoff cutoff method. For direct comparison we therefore set $m=0$ and choose $\mu=1$ in the following subsections. The previous result that we are citing, is the following: 
\begin{eqnarray}
\Tilde{\Gamma}_{{\rm ren}}^{{\rm sp}}(m=0) &\overset{\alpha \to 0}\sim& \frac{1}{12} (\ln \alpha)^2 + \left(-\frac{11}{72} + \half \ln 2 - \frac{\gamma}{6} \right)\ln\alpha\;.  \label{GammasSchubert0}  
\end{eqnarray}
This equation provides the structure of the divergences of the effective action 
as $\alpha \to 0$.
Here, we shall denote the subleading coefficient as $b_{sp}= \left(-\frac{11}{72} + \half \ln 2 - \frac{\gamma}{6} \right)$.

\subsection{WKB-type}
In this section, we use the expressions for the low-modes that were derived by means of the WKB-type method to analyze the structure of the divergent behavior of the one-loop effective action as $\alpha \to 0$ for the massless limit.
 
By considering a massless field, an important simplification occurs in equation (\ref{Seq}) since $W_l (0,r) \equiv \frac{4l+3}{r}$. It is convenient to use the form of (\ref{WKBpartialdet}) that does not involve the mass. Now, the relevant scale of length is $1/\sqrt{\alpha}$. We will omit the indices in $V^{\pm}_{l_3}(r) \equiv V(r)$ and $W_l(0,r) \equiv W(r)$ for simplicity. We have
\bear
S^{l,l_3}_{\pm} (\infty)  &=& \int_0^{\infty} \left( \sqrt{V(r) + \frac{W^2(r)}{4}} - \frac{W(r)}{2}\right) ~dr \label{WKBrvariable} \\ \nn 
&-& \int_0^{\infty} \left( \sqrt{V(r) + \frac{W^2(r)}{4}} - \frac{W(r)}{2}\right) \frac{2 V' + WW'}{(4 V + W^2)^{3/2}}~dr + \cdots
\ear

The first approximation to the partial wave amplitude is given by the first term in (\ref{WKBrvariable}), that is
\be
S_{\pm}^{l,l_3} (\infty) = \int_0^{\infty} \left(\sqrt{\frac{c^2}{4r^2} + V(r) } - \frac{c}{2r}  \right) dr + \cdots \label{WKBmasslessint}
\ee
where $c= 4l+3$. We are interested in obtaining the asymptotic divergent behaviour as $\alpha \to 0$ of this expression. To that end, observe that as $\alpha \to 0$ the large values of $r$ in the integral (\ref{WKBmasslessint}) contribute the most, and by changing variable to $u= \alpha r^2$ the jacobian $1/\sqrt{u}$ factor shifts the relevant region to a neighborhood of $u = 0$ rather than a neighborhood of $u = \infty$, so that classical local analysis techniques can now be used. By the same token, if we are interested only in the leading behavior, it is then appropriate to perform the replacement $e^{-u} \to 1$ in the potential $V(\sqrt{u/\alpha})$, we will call such function $ \alpha \bar{V}(u)$ so that
\be
V(r) \sim \alpha \bar{V}(u)\;,\qquad \alpha \to 0 \;.
\ee
The rational function $\bar{V}(u) = \frac{a_0 + a_1 u + a_2 u^2}{(u+ \alpha)^2}$ approximates $V(u)$, notice that the coefficients $a_i$ are functions of $\alpha$. At this stage, however, it would be incorrect to simply replace $V \to \alpha \bar{V}$ in (\ref{WKBmasslessint}), as that would lead to divergences for $u\to \infty$, this happens because we already eliminated the suppressing exponential decay for large $u$. This last malady can be remedied by introducing an arbitrary constant (meaning $\alpha$-independent) cut-off to regulate the diverging integral, we chose for simplicity $u_{max} =1$, which is effectively equivalent to the $\alpha$-dependent cut-off in $r$ given as $r_{max} = 1/\sqrt{\alpha}$. The following asymptotic estimate is obtained:
\be
\mathcal{S}_0 (\infty) = \int_0^{\infty} \sigma_0 (z)\,dz  \sim \int_0^1 \left(\sqrt{\frac{c^2}{4u} + \bar{V}(u) } - \frac{c}{2\sqrt{u}}\right) \frac{du}{2\sqrt{u}},\qquad \alpha \to 0.
\ee
Thus, the explicit integral is
\be
\mathcal{S}_0(\infty) \sim \int_0^1 \left(\frac{\sqrt{b_0 + b_1 u+ b_2 u^2 + b_3 u^3}}{4u(u+\alpha)}
 - \frac{c}{4u} \right)du\;.
\ee
Now we eliminate the explicit $\alpha$ dependence on the $b_i$'s by noticing that , $b_0 = \alpha^2 c^2$ $b_1/b_0 = \beta_1/\alpha$, $b_2/b_0 = \beta_2 /\alpha^2$, $b_3/b_0 = \beta_3/\alpha^2$ where $\beta_i$'s have finite non-zero values in the limit $\alpha \to 0$, we are led to

\be
\mathcal{S}_0 (\infty) \sim  \frac{c}{4} \int_0^1 \left(\left( \frac{1}{u} - \frac{1}{u+\alpha} \right)\sqrt{1 + \beta_1 (u/\alpha) + \beta_2 (u/\alpha)^2 + \beta_3 u^3 /\alpha^2} -\frac{1}{u}\right) \,du
\ee
Using a Taylor expansion around $u=0$ and the asymptotic relation $\int_0^1 \frac{u^n/\alpha^n}{u+\alpha} \,du\sim (-1)^n \ln \alpha, \,\, \alpha \to 0$, we isolate the relevant contribution, indeed the $\beta_3$ term will not contribute due to the mismatch of powers of $u$ and $\alpha$. We obtain the following asymptotic expression for $\alpha \to 0$:

\bear
 \mathcal{S}(\infty)  \sim && \frac{c}{4}  \left(\int_0^1 \frac{\sqrt{1 + \beta_1 (u/\alpha) + \beta_2 (u/\alpha)^2 } -1}{u}\,du   -  \int_0^1 \frac{ \sqrt{1 + \beta_1 (u/\alpha) + \beta_2 (u/\alpha)^2 }}{u+\alpha} \,du \right) \nn \\
\ear

Both integrals have contributions that cancel out partly and produce a $\ln \alpha$ growth once subtracted. It is now convenient to introduce the new variable $\eta = u/\alpha$ so that for $\alpha \to 0$:

\be
\mathcal{S}_0 (\infty) \sim \frac{c}{4} \left( \int_0^{1/\alpha} \frac{\sqrt{1+ \beta_1 \eta + \beta_2 \eta^2} -1}{\eta}\,d\eta   - \int_0^{1/\alpha}  \frac{\sqrt{1+ \beta_1 \eta + \beta_2 \eta^2}}{\eta+1} \,d\eta \right)
\ee
Notice how it is now the upper limit region $\eta \approx 1/\alpha$ that is most important (corresponding to $u=1$, i.e. $r = 1/\sqrt{\alpha}$) and this expression can be evaluated straightforwardly by Taylor expansion around $\eta =\infty$, resulting in
\be
\mathcal{S}_0(\infty) \sim \frac{1}{4} \left( c- \sqrt{c^2 + 16 l_3 + 4 +8s } \right) \ln \alpha
\ee
where $s = \pm 1$ represents the polarization of the partial-waves as indicated in (\ref{GammaPartialwaves}). 

At this level of approximation, the low modes contribution to the spinor effective action can be obtained from (\ref{GammaGYT}), yielding:
 \be  \label{LowModesWKBmassless}
\Tilde{\Gamma}_{{\rm L}}^{(0)} \sim - \sum_{l=0,\half,\cdots}^{L} (2l+1) \sum_{l_3 = -l}^{l}  \sum_{s = \pm}\frac{1}{4} \left( c- \sqrt{c^2 + 16 l_3 +4 + 8s} \right) \ln \alpha
\ee 
In order to evaluate the r.h.s. of (\ref{LowModesWKBmassless}) one should notice that the first term gives a closed form  
\[
\sum_{l=0,\half,\cdots}^{L}  \sum_{l_3 = -l}^{l}  \sum_{s = \pm}(2l+1) c = 
\frac{2}{3} (1 + L) (1 + 2 L) (9 + 22 L + 12 L^2)
\]
whereas the second term can be evaluated by an Euler-Maclaurin expansion for large $L$ to the third order in both $l_3$ and $l$ keeping terms up to order $\mathcal{O}(1/L)$. To this end it is useful to notice the following relation

\be
\sum_{l_3=-l}^{l} \sqrt{c^2 + 16 l_3 +4 + 8s} = 4 \left(\zeta_{\rm{H}} (-1/2, \omega)  -\zeta_{\rm{H}} (-1/2, \omega + 2l+1)  \right)
\ee
where $\omega = (13 + 8 l + 16 l^2 + 8 s)/16$ and  $\zeta_{\rm{H}} (\lambda,z) = \sum_{n=0}^{\infty} 1/(n+z)^{\lambda}$ is the Hurwitz zeta function. After completing the calculation one finds

\be 
\Tilde{\Gamma}_{{\rm L}}^{(0)} \sim \left( \frac{2}{3} L^2 +L + k - \frac{11}{24}\ln L \right)\ln\alpha  + \mathcal{O} (1/L) \; \ln \alpha
\ee
where
\bear                     
k& =& \frac{31489}{72000} + \sqrt{\frac{3}{7}} \frac{9610449}{38416000} - \frac{516233}{
 2000000 \sqrt{5}} - \frac{\rm{arccsch}\,2}{192} - \frac{265 \ln 2}{192} + 
 \frac{23}{64} \ln(5 + \sqrt{5}) \nn \\
 &+& \frac{23}{192} \ln(1 + \sqrt{21}) - 
 \frac{\ln(5 + \sqrt{21})}{64} = 1.63365\cdots
\ear

and after adding the contribution of the high modes (also in asymptotic form), we obtain the following expression the effective action:

\be \label{wkbspinorzeromass}
\Tilde{\Gamma}^{(0)} \sim \frac{(\ln \alpha)^2}{12} + \left(1.83803\cdots - \frac{1}{8} \ln L  +\mathcal{O}(1/L) \right) \ln \alpha
\ee
From this last expression we remark the following: First, that we have indeed obtained the correct structure of the small $\alpha$ divergences, identifying that the $(\ln \alpha)^2$ term is entirely derived from the high-modes while the subleading $\ln \alpha$ behavior comes from both types of modes. One should also notice that the $\ln \alpha$ contribution resulting from the low-modes is a non-trivial result. Given the fact that the partial-wave-cutoff method requires delicate cancellations from both kinds of modes, it is also remarkable that the quadratic and linear terms in $L$ are exactly cancelled out in (\ref{wkbspinorzeromass}), with only the $\ln L$ divergence left. Unfortunately, this means that the performing the second approximation (small $\alpha$) results in information loss that prevents the use of the full partial-wave-cutoff method for achieving an even simpler expression for the renormalized effective action.

The calculation derived from the next WKB type order of approximation, namely $\mathcal{S}_1$, is more involved, but by following the exact same procedure outlined above we find

\bear
\mathcal{S}_1 \sim &-& \frac{c}{4}\int_0^1 \Bigg( \left( \frac{1}{u} - \frac{1}{u+\alpha} \right) \sqrt{1 + \beta_1 (u/\alpha) + \beta_2 (u/\alpha)^2 + \beta_3 u^3 /\alpha^2 } \nn \\  &\times & \sum_{p=0}^{\infty} Y_p (u/\alpha)^p  -\frac{1}{u} \Bigg)
\,du
\ear
where the coefficients have a growth rate $Y_p \sim O(l_3^p /l^{2p+1})$. We are interested only in precision up to order $\mathcal{O} (\ln L)$ in the effective action so that terms of order $\mathcal{O} (1/L)$ and smaller play no part in the low modes contribution, and hence we can truncate the series at $p=2$. Explicitly 

\be
Y_0 =-\frac{1}{2c},\,\,\, Y_1 = \frac{12(l_3 + s)}{c^3},\,\,\,
Y_2 = \frac{240 (l_3 + s)^2 + c^2 ( 20 l_3 + 30s -5)}{c^5}
\ee
 
Here we encounter poles of first and second order together with the logarithmic divergence in $\alpha$, namely

\begin{eqnarray}
\mathcal{S}_1 (\infty) & \sim & \frac{   (1-2\beta_2)Y_0 - 4 \beta_2 Y_1 (\beta_1 - 2 \beta_2)  - 
   Y_2 (\beta_1^2 + 4 \beta_1 \beta_2 - 
      4 \beta_2 (1 + 2 \beta_2)))}{8 \beta_2^{3/2}} \ln \alpha \nn \\
      &+&\frac{1}{\alpha} \left( \frac{Y_2 \beta_1 + 2\beta_2 (Y_1 - Y_2)}{2\sqrt{\beta_2}} -Y_1 \right) + \frac{Y_2(\sqrt{\beta_2} -1)}{2} \frac{1}{\alpha^2}
\end{eqnarray}
The presence of poles signals the breakdown of the WKB-type subleading order contribution when a second approximation, namely small $\alpha$, is taken. And here we also loose information about the $L$ behaviour, a situation that may be expected from the fact that the $\alpha \to 0$ limit is ill behaved.

\subsection{Dominant Balance }

In this subsection, we investigate further the asymptotic prediction for the leading and subleading terms in the massless case for $\alpha \to 0$, but this time, by using the method of dominant balance. 

Our starting point is the massless limit of equation (\ref{Sigma0}):
\begin{equation}
  \tilde{\Sigma}_0 (r) = \frac{1}{r^{4l+3}} \int_0^r V(\rho) \rho^{4l+3} d\rho  
\end{equation}
where we keep using the shorthand $V_{l_3}^{\pm}  (r)= V (r)$. It will be convenient to change the integration variable to $u= \rho/r$, obtaining

\begin{equation} \label{Sig0}
   \tilde{\Sigma}_0 (r)= r \int_0^{1} V(ur) u^{4l+3} du
\end{equation}
This leads to the following expression for the partial-waves:
\be
\tilde{S}_0 (\infty) = \int_0^{1} du\, u^{4l+3} \int_0^{\infty} dr \,V(ur) r
\ee
Now we change variables again to $\bar{r}=ur$ and are left with
\begin{equation}
    \tilde{S}_0 (\infty) = \frac{1}{4l+2} \int_0^{\infty} V(\bar{r}) \bar{r}\, d\bar{r}
\end{equation}
Then, the low-modes contribution to the effective action for this order of approximation is
\begin{eqnarray}
\Tilde{\Gamma}_{{\rm L}}^{(0)} &=& - \sum_{l = 0, 1/2, \cdots}^L (2l+1)\sum_{l_3 = -l}^l \sum_{s=\pm} \tilde{S}_0 (\infty) \nn \\ 
&=&-\sum_{l = 0, 1/2, \cdots}^L (2l+1)\sum_{s=\pm}\sum_{l_3= -l}^l \frac{1}{4l+2} \int_0^{\infty} V(r) r dr \\
&=& -\half \sum_{l = 0, 1/2, \cdots}^L (2l+1) \sum_{s=\pm} \int_0^{\infty} V(r) r \,dr \nn
\end{eqnarray}
and after calculating by the same procedure outlined in the previous section we obtain the following asymptotic relation

\be
\Tilde{\Gamma}_{{\rm L}}^{(0)} \sim  \left( L^2 + \frac{3}{2}L + \half \right)\ln \alpha
,\quad \alpha\to 0 \ee
and, after adding the high modes contribution $\Tilde{\Gamma}^{(0)}= \Tilde{\Gamma}_{{\rm L}}^{(0)} + \Tilde{\Gamma}_{{\rm{H, ren}}}$ we obtain for the effective action

\be
\Tilde{\Gamma}^{(0)} \sim \frac{1}{12} (\ln \alpha)^2  +\left(\frac{11}{360} +\frac{\gamma}{6} + \frac{5}{6} \ln 2 \right) \ln \alpha +  \mathcal{O}(L^2) \ln \alpha  , \quad \alpha \to 0.
\ee
where $\gamma$ is the Euler-Mascheroni constant.

Notice that we have once again obtained the correct divergence structure with the correct leading term. However, this time we also get the same arithmetic structure for the $L$-independent part subleading coefficient as in (\ref{GammasSchubert0}), namely $r_1 + r_2 \ln2 + r_3 \gamma $ where $r_{1,2,3} \in \mathbb{Q}$. This is remarkable given that the methods used in each case are very different. For comparison:

\begin{eqnarray}
b_{sp}^{(0)} &=& \frac{11}{360} +\frac{\gamma}{6} 
+ \frac{5}{6} \ln 2= 0.704381\cdots \\
b_{sp} &=& -\frac{11}{72} - \frac{\gamma}{6}+ \half \ln 2  =0.0975932\cdots
\end{eqnarray}

One should notice the fact that we have obtained an imperfect cancellation of the quadratic, linear and logarithmic terms in $L$ that accompany the subleading term $\ln \alpha$, this is a consequence of the small $\alpha$ approximation, and suggests that the renormalization procedure, as carried out by means of an angular momentum cutoff, is not completely consistent with the two approximations in effect here.

We have also calculated the next order of approximation. To that end consider

\be
\tilde{\Sigma}_1 (r) = -\frac{1}{r^{4l+3}} \int_0^r \tilde{\Sigma}_0^2(\rho) \rho^{4l+3} d\rho
\ee
changing variable to $\rho = ur$ one gets

\begin{equation} \label{Sig1}
 \tilde{\Sigma}_1 (r) = - r \int_0^{1} \tilde{\Sigma}_0^2 (ur) u^{4l+3}\, du
\end{equation}
and after integrating over $r$ we obtain for the partial wave

\begin{equation}
    \tilde{S}_1 (\infty) = -\int_0^1 du\, u^{4l+3} \int_0^{\infty} dr \, r  \tilde{\Sigma}_0^2 (ur)
\end{equation}
changing variable to $\rho = ur$

\begin{eqnarray}
 \tilde{S}_1 (\infty)& = & - \int_0^{1} du\, u^{4l+1} \int_0^{\infty}\rho \nn \tilde{\Sigma}_0^{2} (\rho) d\rho \\ 
& =& - \frac{1}{4l+2} \int_0^{\infty}\rho \tilde{\Sigma}_0^{2} (\rho) d\rho\\ \nn
&=&- \frac{1}{4l+2} \int_0^{\infty}\rho \left(\rho \int_0^1 V(u\rho) u^{4l+3}\,du \right)^2 d\rho
\end{eqnarray}
After renaming $\rho \to r$ and the second variable $u\to w$ we finally arrive at 

\begin{equation}
\tilde{S}_1 (\infty) = -\frac{1}{4l+2} \int_0^{1} \int_0^1 du dw \, (uw)^{4l+3} \int_0^{\infty} r^3 V(ur) V(wr)\, dr
\end{equation}
Then low-modes contribution to the effective action is

\begin{eqnarray}
\Tilde{\Gamma}_{\rm L}^{(1)} &=& \Tilde{\Gamma}_{\rm L}^{(0)} \nn \\ &+& \half \sum_{l=0,1/2,\cdots}^{L} \sum_{s= \pm} \sum_{l_3 = -l}^{l} \int_0^{1} \int_0^1 du dw \, (uw)^{4l+3} \int_0^{\infty} r^3 V(ur) V(wr)\, dr
\end{eqnarray}
This expression can be analyzed as before and its asymptotic limit $\alpha \to 0$ can readily be obtained as in the previous approximation level. Overall, the next order of approximation of the low modes contribution to the effective action is

\begin{eqnarray}
\Tilde{\Gamma}^{(1)}_{\rm L}  &\sim& \Tilde{\Gamma}_{\rm L}^{(0)} - \left(\frac{L^2}{3} + \frac{L}{2} + \left(\frac{1}{6} + \frac{1}{24}\gamma + \frac{11}{24}\ln 2 \right) + \frac{11}{24}\ln L  \right) \ln \alpha 
\end{eqnarray}
Once the high-modes contribution is accounted for we obtain 

\be
\Tilde{\Gamma}^{ (1)} \sim  \frac{1}{12}(\ln \alpha)^2 +\left(-\frac{49}{360} + \frac{1}{8}\gamma + \frac{3}{8} \ln 2 \right)\ln \alpha + \mathcal{O}(L^2) \ln \alpha
\ee
which improves the prediction for the subleading coefficient to 

\[
b_{sp}^{(1)} =-\frac{49}{360} + \frac{1}{8}\gamma + \frac{3}{8} \ln 2  = 0.195971\cdots
\]
Even though we can't exactly obtain the predicted value by this means, we can appreciate two things, first that the non-trivial algebraic structure of the coefficient $b_{sp}$ is reproduced, and second that as we went from the first $b_{sp}^{(0)}$, to the second order, $b_{sp}^{(1)}$, of approximation the predicted coefficient gets closer to $b_{sp}$. We take this as an indication that a better result may be obtained, including a better behaviour of the still remaining divergent terms in $L$, if higher orders of approximation are included.

\section{Conclusions  \label{conclusions}}

We have investigated the use of classical asymptotic methods to obtain quadrature expressions for determinants through the GYT. We specifically applied this ideas to the calculation of the QED one-loop effective action under background field with an $O(2)\times O(3)$ symmetry but including an arbitrary profile function $g(r)$. We improved on previous work on the subject by providing useful asymptotic expressions for the solution of the low-modes equation $(\ref{Seq})$ which so far, had only been solved numerically. The approach we propose classifies the types of response that the field has in different, complementary, regimes according to the background properties. These two regimes are termed the DM and WKB-type method. The asymptotic methods presented for each regime are tested against the exact solution of the problem, showing good precision under different  parameters such as field-intensity, falloff rate and different mass regimes. 

We have demonstrated that both methods have enough precision to allow for the calculation of the full one-loop effective action using the partial-wave cutoff method; this demands high numerical accuracy since the result is produced by two contributions (low-modes and high-modes) with cancelling divergences. 

Both methods  can be systematically developed to higher orders although the work involved may become cumbersome. The main results that are shown in equations (\ref{Sseries}) and (\ref{WKBpartialdet}) are given in terms of the arbitrary background profile function $g(r)$ and may be useful for both spinor and scalar cases. 

The way these two approximations work is different to any of the previously available approximation methods as they are not directly related to the small or large mass regimes, and they are not similar to the derivative expansion either. 
While it was observed that the dominant balance method, when applied to the calculation of the effective action, requires a computation time that is similar to that of the old numerical method, the WKB-type method is much faster.

For a specific profile function such as (\ref{gdef}) the expressions for the low-modes derived from the asymptotic approximations may be analyzed to extract information such as possible divergences with respect to some parameters. Remarkably, for the massless case, and approximating for small $\alpha$ the sum of the low-modes can be done analytically.

In the
case presented, it was shown that although we are limited by starting already with an approximation, the correct non-trivial leading structure of the divergences of $\Tilde{\Gamma}_{{\rm ren}}(m=0)$ as $\alpha \to 0$ was obtained from asymptotics. To be specific, it was found that the leading divergent behavior comes only from the high-modes contribution, while the subleading behavior comes from both the high and low modes contributions. As expected, it was confirmed that the asymptotic expansion of the low-modes contribution in small $\alpha$ contributions looses information on the coefficients of $\ln L$, $L$ and $L^2$, making that approximation incomplete to be used for full calculations of the effective action by the partial-wave-cutoff method. 
However, we must recall that the numerical evidence (as in figure \ref{figure6}) shows that the effective action can indeed be calculated if one uses the complete $\alpha$ dependence.

The subleading behaviour was investigated in the $\alpha \to 0$ massless regime, where we were able to derive an approximation to the result reported in \cite{PhysRevD.87.125020}, in particular recovering the exact same arithmetic structure over the field of rationals.

The question remains as to what extent the asymptotic methods presented here can be used in combination to the partial-wave-cutoff method for broader analytical purposes. So far, the previously developed methods to calculate the high-mode contribution allow for dimensional renormalization. But it is possible that a different renormalization scheme could result even more adequate. Further research along these veins is still needed to settle this question.

\section*{Acknowledgements}
We would like to thank C. Schubert and J. P. Edwards for their useful comments. A. H. received funding from  PRODEP, grant number 511-6/2020-8588.

\appendix

\newpage
\section{The high-modes contribution\label{highmodes} }

For large values of $L$, the sum of the highmodes is well approximated by the following expression:

\begin{equation*}
\Gamma_{{\rm{H}}}^{\rm{ren}} = \int_0^\infty dr  \left( Q_{{\rm {log}}}(r)\ln L + \sum_{n = 0}^2 Q_n ( r ) L^n  +  \sum_{n = 1}^N Q_{-n}( r ) \frac{1 }{ L^n }  \right) + O(\frac{1}{L^{n+1}}) \, .
\label{gammaren}
\end{equation*}
For the spinor case we have:

\begin{align}
Q_2(r) &=  \frac{8 \, g(r)^2 r^3}{3 \sqrt{\tilde{r}^2 + 4}} \nn \\
Q_1(r) &= \frac{2 r^3 (3 \tilde{r}^3 + 8) g(r)^2}{(
\tilde{r}^2 + 4)^{3/2}} \nn \\
Q_{{\rm {log}}}(r) &= -\frac{1}{6}\, r^3 (20 g(r)^2 + 10 g(r)g'(r) r + g'(r)^2
r^2 ) \nn \\
Q_0(r) &= \frac{r^3}{45 ( \tilde{r}^2 +
4)^{7/2}} \Bigg[ -6 r^4 (5 \tilde{r}^4 + 28 \tilde{r}^2 + 32 )g(r)^4
\nn \\ &+ 15 (33 \tilde{r}^6 +  335 \tilde{r}^4  + 1192 \tilde{r}^2 + 1600) g(r)^2  \nn \\
&+ 10 r (15 \tilde{r}^6 + 184 \tilde{r}^4 + 776 \tilde{r}^2 + 1120) g(r)g'(r) \nn \\
&+  5 r^2 (3 \tilde{r}^6 + 38 \tilde{r}^4+ 160 \tilde{r}^2 + 224) g'(r)^2 + 20 r^2 (4 + \tilde{r}^2)^2
g(r)g''(r)  \Bigg]
 \nn \\
&- Q_{{\rm {log}}}(r)   \ln \Bigg( \frac{ \mu r}{2 + \sqrt{\tilde{r}^2 + 4}}
\Bigg)
 \nn \\
Q_{(-1)}(r) &= -\frac{r^3}{360 ( \tilde{r}^2 + 4)^{9/2}} \Bigg[ 540
r^4( \tilde{r}^6 + 4 \tilde{r}^4 )g(r)^4 \nn\\
&+180 r^2(  \tilde{r}^6 + 16 \tilde{r}^4 + 80\tilde{r}^2 +
 128)g'(r)^2 \nn  \\
&+ (-359 \tilde{r}^8 + 8008 \tilde{r}^6 + 99336 \tilde{r}^4 + +311040 \tilde{r}^2 +460800) g(r)^2 \nn \\
&+1440 r( 2 \tilde{r}^6 + 21 \tilde{r}^4 + 92
\tilde{r}^2 + 160)g(r) g'(r) \nn \\
&-360 r^2(\tilde{r}^6 + 8 \tilde{r}^4 + 16
\tilde{r}^2 )g(r) g''(r)
 \Bigg] \, ,     \label{spinorQs}
\end{align}
where $\tilde{r} = mr/L$. 

The expression for $\Gamma_{{\rm{H}}}^{\rm{ren}}$ contains $L^2$, $L$ and
$\ln (L)$ terms  that diverge as $L \to \infty$, but these divergences
exactly cancel those of the $\Gamma_{{\rm{L}}}$ contribution. 

The corresponding expressions for the scalar case are:
\begin{align}
Q_2(r) &=  -\frac{8 \, g(r)^2 r^3}{3 \sqrt{\tilde{r}^2 + 4}} \nn \\
Q_1(r) &= -\frac{2 r^3 (3 \tilde{r}^3 + 8) g(r)^2}{(
\tilde{r}^2 + 4)^{3/2}} \nn \\
Q_{{\rm {log}}}(r) &= -\frac{1}{12}\, r^3 (8 g(r)^2 + 4 g(r)g'(r) r + g'(r)^2
r^2 ) \nn \\
Q_0(r) &= \frac{r^3}{90 ( \tilde{r}^2 +
4)^{7/2}} \Bigg[ 12 r^4 (5 \tilde{r}^4 + 28 \tilde{r}^2 + 32 )g(r)^4
\nn \\ 
& -30 (9 \tilde{r}^6 +  47 \tilde{r}^4  + 40 \tilde{r}^2 + 64) g(r)^2  \nn \\
&+ 20 r (3 \tilde{r}^6 + 32 \tilde{r}^4 + 88 \tilde{r}^2 + 32) g(r)g'(r) \nn \\
&+  5 r^2 (3 \tilde{r}^6 + 32 \tilde{r}^4+ 112 \tilde{r}^2 + 128) g'(r)^2 - 40 r^2 (4 + \tilde{r}^2)^2
g(r)g''(r)  \Bigg]
 \nn \\
&- Q_{{\rm {log}}}(r)   \ln \Bigg( \frac{ \mu r}{2 + \sqrt{\tilde{r}^2 + 4}}
\Bigg)
 \nn \\
Q_{(-1)}(r) &= \frac{r^3}{4 ( \tilde{r}^2 + 4)^{9/2}} \Bigg[ 6
r^4 \tilde{r}^4 ( \tilde{r}^2 + 4  )g(r)^4 \nn\\
&- ( 4 \tilde{r}^8 + 7 \tilde{r}^6 + 48 \tilde{r}^4 + 1152 \tilde{r}^2 +
 1024)g(r)^2 \nn  \\
&-16 r( \tilde{r}^6 + 15 \tilde{r}^4 + 
52 \tilde{r}^2 + 32)g(r) g'(r) \nn \\
&-4 r^2( \tilde{r}^2 + 4  )^2 \{ ( \tilde{r}^2 + 2  )g'(r)^2   + 
\tilde{r}^2 )g(r) g''(r) \}
 \Bigg] \, .     \label{scalarQs}
\end{align}

\section{Manipulations} \label{A}

In this appendix we show the transformations necessary to arrive at the given expressions already stated. To this end consider the following identity

\be \label{derbessel}
-\frac{d}{dx} \left( \frac{K_{\mu} (x)}{I_{\mu} (x)} \right) = \frac{1}{x I_{\mu}^2 (x)}, \quad \mu \in \mathbb{Z} \qquad
\ee
which allows us to rewrite (\ref{s0first}), as the double integral

\be
S_{0} (z) = -\int_0^z \int_0^z dx ~dy~ \theta (x-y) \frac{d}{dx}  \left(\frac{K_{2l+1} (x)}{I_{2l+1} (x)} \right)  y I_{2l+1}^2 (y)   v_{l_3}(y)
\ee
integrating by parts with respect to $x$ eliminates one integration, namely

\begin{eqnarray} \nonumber
S_0 (z) &=& -\int_0^z \int_0^z dx ~dy~ \delta (x-y) \frac{K_{2l+1}(x)}{I_{2l+1} (x)} y I_{2l+1}^2 (y) v(y)  \\
&+& \int_{0}^z dy~ \frac{K_{2l+1} (z)}{I_{2l+1} (z)} y I_{2l+1}^2 (y) v(y)
\end{eqnarray}
from which

\be
S_{0}(z) = \int_0^{z} dy ~v(y) y \frac{I_{2l+1} (y)}{I_{2l+1}(z)}   \left(K_{2l+1}(y) I_{2l+1}(z) -K_{2l+1}(z) I_{2l+1}(y) \right)
\ee
follows readily. For $S_1 (z)$ we shall use the identity

\be
-\half \frac{d}{dy} \left(\frac{K_{\mu}(y)}{I_{\mu}(y)}\right)^2 = \frac{K_{\mu} (y)}{y I_{\mu}^3 (y)}, \quad \mu \in \mathbb{Z}
\ee
Starting from 

\bear
S_1 (z) &=& - \int_0^z ds~ \frac{1}{s I_{2l+1}^2(s)} \int_0^s dx~ \frac{1}{x I_{2l+1}^2 (x)} \left( \int_0^{x} dy~ I_{2l+1}^2 (y) v(y) y \right)^2  \nn \\ \nn
&=&  \int_0^z dx~ \int_0^z ds~ \theta (s-x) \left( \frac{d}{ds} \frac{K_{2l+1}(s)}{I_{2l+1}(s)} \right) \frac{1}{x I_{2l+1}^2 (x)}\left( \int_0^{x} dy~ I_{2l+1}^2 (y) v(y) y \right)^2 
\ear
Integrating by parts with respect to $s$ one obtains

\bear
S_1 (z) =\frac{K_{\mu}(z)}{I_{\mu}(z)} \int_0^{z} dx~ \frac{1}{xI_{2l+1}^2 (x)}
\left( \int_0^{x} dy~ I_{2l+1}^2 (y) v(y) y \right)^2 
-\int_0^z dx~ \frac{K_{\mu}(x)}{x I^3_{\mu}(x)} \left( \int_0^{x} dy~ I_{2l+1}^2 (y) v(y) y \right)^2
\ear
and may be recast into the form

\bear
S_1 (z) = - \frac{K_{2l+1}(z)}{I_{2l+1}(z)} \int_0^z dx~ \left(\frac{d}{dx}\frac{K_{2l+1}(x)}{I_{2l+1}(x)}\right)
\left( \int_0^{x} dy~ I_{2l+1}^2 (y) v(y) y \right)^2 \label{identS1} \\ \nn
+ \half \int_0^z dx~ \frac{d}{dx}\left( \frac{K_{2l+1}(x)}{I_{2l+1}(x)}\right)^2 \left( \int_0^{x} dy~ I_{2l+1}^2 (y) v(y) y \right)^2 \nn
\ear
the first term on the right can also be written as 

\be
- \frac{K_{2l+1}(z)}{I_{2l+1}(z)} \int_0^z dx\int_0^{z} dy \int_0^{z} dt~ \left(\frac{d}{dx}\frac{K_{2l+1}(x)}{I_{2l+1}(x)}\right)  \times \theta (x-t) \theta (x-y) I_{2l+1}^2 (y) v(y) y I_{2l+1}^2 (t) v(t) t 
\ee

after integration by parts with respect to $x$ this becomes 

\bear
- \left(\frac{K_{2l+1}(z)}{I_{2l+1}(z)}\right)^2 \left(\int_0^z dy~ I_{2l+1}^2 (y) v(y) y \right)^2 + 2 \frac{K_{2l+1}(z)}{I_{2l+1}(z)} \int_0^z dt~ K_{2l+1}(y)I_{2l+1}(y) v(y) y \int_0^y dt~ I_{2l+1}^2 (t) v(t) t
\ear
the remaining term in (\ref{identS1}) can be likewise be written as 

\[
\frac{1}{2} \int_0^z dx \int_0^z dy \int_0^z dt~ \theta (x-t) \theta (x-y) \frac{d}{dx} \left(\frac{K_{2l+1}(x)}{I_{2l+1}(x)} \right)^2
\]
which after integration by parts with respect to $x$ results into

\[
  \half \left(\frac{K_{2l+1}(z)}{I_{2l+1}(z)} \right)^2 \left( \int_0^{z} dy ~I_{2l+1}^2(y) v(y) y \right)^2 
 -  \int_0^z dt ~K_{2l+1}^2 (t) v(t) t 
\int_0^t dy ~I_{2l+1}^2 (y) v(y) y  
\]
overall we have obtained (\ref{S1}). When $z \to \infty$ we obtain the following values

\bear
S_0 (\infty)&=& \int_{0}^{\infty}dy~ K_{2l+1}(y) I_{2l+1} (y) v(y) y  \\
S_1 (\infty)&=& - \int_0^\infty dy~ K^2_{2l+1} (y) v(y) y \int_0^y dx~ I_{2l+1} (x) v(x) x \nn
\ear
this approximation is good whenever $|\Sigma_0 (z)|^2 \ll 1$, and this is guaranteed for instance when $\max_{z \in \mathbb{R}_+} |v(z)/w^2(z)| \ll 1$, since in such case the maximum (given by $\Sigma_0 ' (z_0) =0$) is small compared to one, say of the order $10^{-1}$. All integrals above can be shown to converge easily for our family of potentials.

\section{Integrals}

In this appendix we collect the dominant asymptotic behaviour as $\alpha \to 0$ of the  integrals needed to find the asymptotic behaviour of the high-modes contribution for the renormalized massless effective action in the limit $\alpha \to 0$. We based our calculation in the lengthy expressions found in \ref{highmodes}.

\begin{equation}
\Gamma_{\rm{H}} =  I_{Q_{log}} \ln L + \sum_{k=0}^2 I_{Q_{k}}L^k + \mathcal{O}(1/L)
\end{equation}
 Where $I_{Q_A} := \int_0^{\infty} Q_A (r) dr$ and the profile function is  $g(r) = \frac{e^{-\alpha r^2}}{1+r^2}$; the integrals in question are evaluated asymptotically

\begin{eqnarray}
\int_0^\infty r^3 g^2 dr &\sim & -\half \ln \alpha \\
\int_0^\infty r^7 g^4 dr &\sim & -\half \ln \alpha\\
\int_0^\infty r^4 g g' dr &\sim &  \ln \alpha\\
\int_0^\infty r^5 g'^2 dr &\sim & -2 \ln \alpha \\
\int_0^\infty r^5 g g'' dr &\sim & -3 \ln \alpha \\
\int_0^\infty r^3 g^2 \ln (r/4) dr &\sim & \frac{1}{8} (\ln \alpha)^2 + \left(\frac{\gamma}{4} + \frac{5}{4}\ln 2\right)\ln \alpha \\
\int_0^\infty r^4 g g' \ln (r/4) dr &\sim & -\frac{1}{4} (\ln \alpha)^2 + \left(\frac{1}{4} - \frac{\gamma}{2}-\frac{5}{2}\ln 2 \right)\ln \alpha \\
\int_0^\infty r^5 g'^2 \ln (r/4) dr &\sim & \half (\ln \alpha)^2 + \left(   5\ln 2 + \gamma - \frac{5}{4} \right)\ln \alpha
\end{eqnarray}
where various degrees of subdominant terms have bee omitted. Finally, the following asymptotic form is found 

 \begin{eqnarray}
 \Gamma_{\rm{H}}^{\rm sp} & \sim & \frac{(\ln \alpha)^2}{12} +\left( -\frac{169}{360} + \frac{\gamma}{6} +\frac{5}{6}\ln 2  -  L -   \frac{2}{3} L^2  +\frac{1}{3} \ln L  + \mathcal{O} (1/L) \right)\ln \alpha  \nn \\   
\end{eqnarray}
For completeness, we give also the result for the scalar case
\begin{eqnarray}
\Gamma_{\rm{H}}^{\rm sc} & \sim & \frac{ (\ln \alpha)^2 }{24}+ \left(  \frac{\gamma}{12} + \frac{113}{720} + \frac{5}{12} \ln 2   + L   + \frac{2}{3} L^2 +  \frac{1}{6}\ln L + \mathcal{O} (1/L) \right) \ln \alpha \nn \\
\end{eqnarray}

\bibliography{Source}

\end{document}